\documentclass[12pt,preprint]{aastex}

\shortauthors{Matthews et al.}
\shorttitle{Feature Movie of Emission 20-100~AU from \SI}

\begin{document}

\newcommand{\ang}{\rm \AA}
\newcommand{\msun}{M$_\odot$}
\newcommand{\lsun}{L$_\odot$}
\newcommand{\days}{$d$}
\newcommand{\degree}{$^\circ$}
\newcommand{\ud}{{\rm d}}
\newcommand{\as}[2]{$#1''\,\hspace{-1.7mm}.\hspace{.0mm}#2$}
\newcommand{\am}[2]{$#1'\,\hspace{-1.7mm}.\hspace{.0mm}#2$}
\newcommand{\ad}[2]{$#1^{\circ}\,\hspace{-1.7mm}.\hspace{.0mm}#2$}
\newcommand{\lsim}{~\rlap{$<$}{\lower 1.0ex\hbox{$\sim$}}}
\newcommand{\gsim}{~\rlap{$>$}{\lower 1.0ex\hbox{$\sim$}}}
\newcommand{\HA}{H$\alpha$}
\newcommand{\HII}{\mbox{H\,{\sc ii}}}
\newcommand{\kms}{\mbox{km s$^{-1}$}}
\newcommand{\HI}{\mbox{H\,{\sc i}}}
\newcommand{\SI}{\mbox{Source~I}}
\newcommand{\SIc}{\mbox{SOURCE~I}}
\newcommand{\jks}{Jy~km~s$^{-1}$}

\title{A Feature Movie of SiO Emission 20-100~AU from 
the Massive Young Stellar Object Orion Source~I}

\author{L. D. Matthews\altaffilmark{1,2},
  L. J. Greenhill\altaffilmark{1}, C. Goddi\altaffilmark{1},
  C. J. Chandler\altaffilmark{3}, E. M. L. Humphreys\altaffilmark{1},
  M. W. Kunz\altaffilmark{3,4}}

\altaffiltext{1}{Harvard-Smithsonian Center for Astrophysics,
60 Garden Street, Cambridge, MA, USA 02138}
\altaffiltext{2}{MIT Haystack Observatory, Off Route 40, Westford, MA USA
01886}
\altaffiltext{3}{National Radio Astronomy Observatory, 
P.O. Box O, Socorro, NM, USA 87801}
\altaffiltext{4}{University of Illinois at Urbana-Champaign,
  Department of Physics, 1110 West Green Street, Urbana, IL 61801}

\begin{abstract}
We present multi-epoch Very Long Baseline Array (VLBA) 
imaging of the 
$^{28}$SiO $v$=1 and $v$=2, $J$=1-0 maser emission toward the
massive young stellar object (YSO) Orion \SI.
Both SiO transitions
were observed simultaneously with an angular resolution of $\sim$0.5~mas 
($\sim$0.2~AU for $d$=414~pc) and a spectral resolution of 
$\sim$0.2~\kms. 
Here we explore the global properties and kinematics of the emission 
through two
19-epoch animated movies spanning 21 months
(2001 March 19 to 2002 December 10). These movies provide 
the most detailed view to date 
of the dynamics and temporal evolution of molecular material
within $\sim$20-100~AU of a massive ($\gsim8M_{\odot}$) YSO. 
As in previous studies, we find that 
the bulk of the SiO masers surrounding \SI\ 
lie in an {\sf X}-shaped locus; the emission
in the South and East arms is predominantly blueshifted and emission in the
North and West is predominantly redshifted. 
In addition, bridges of intermediate-velocity emission are
observed connecting the red and blue sides of the emission distribution.
We have measured proper motions of over 1000 individual maser
features and find that these motions  
are characterized by a combination of 
radially outward migrations along the four main maser-emitting 
arms and motions tangent to the intermediate-velocity 
bridges. 
We interpret the SiO masers as arising from a wide-angle bipolar wind emanating
from a rotating, edge-on disk. The detection of maser features along
extended, curved filaments suggests that magnetic fields may 
play a role in launching and/or shaping the wind.
Our observations appear to support a picture in which
stars with masses as high as at least $8M_{\odot}$ form via
disk-mediated accretion. However, we cannot yet rule out that the \SI\ disk
may have been formed or altered following a recent close encounter.

\end{abstract}


\section{Introduction}
At a distance of $\sim414$~pc (Menten et al. 2007; Kim et al. 2008), radio
``\SI'' in the Kleinmann-Low (KL) nebula of Orion is believed to
be the nearest example of a massive young stellar object (YSO). 
\SI\ is highly embedded and has no
optical or infrared counterpart;  at wavelengths of 8 and 22$\mu$m 
Greenhill et al. (2004a) estimated continuum optical depths of $>$300. 
However, high-resolution 7-mm radio continuum observations have
revealed an extended ($\sim$70~AU across) source that Reid et al.
(2007) proposed may be an ionized accretion disk surrounding
an early B-type star. 

In addition to its proximity, Orion~KL 
has the distinction of being one of only
three star-forming regions known to exhibit SiO maser emission
(Hasegawa et al. 1986; Zapata et al. 2009), and these masers were 
definitively linked with \SI\ by Menten \& Reid (1995).
SiO masers offer
key advantages for studying stellar sources in that they
arise at small radii
($T_{\rm ex}>10^{3}$~K, $n({\rm H}_{2})\sim10^{10\pm1}$~cm$^{-3}$),
are unaffected by
extinction from dust and neutral gas, and can sample gas kinematics
via emission line Doppler shifts and proper motions. 
Moreover, their high brightness temperature makes them observable with 
extraordinarily high angular
resolution using Very Long Baseline Interferometry (VLBI). Indeed,
previous VLBI observations have established that
the SiO masers 
lie at projected distances of $\sim$20-100~AU from \SI\
(Greenhill et al. 1998, 2004b; Doeleman et al. 1999, 2004; 
Kim et al. 2008). These scales
are of considerable interest, since they 
correspond to regions where accretion is expected to occur and where winds 
and outflows from YSOs are expected to be
launched and collimated. At present, little is known about the dynamics
of these regions in higher mass YSOs ($M_{\star}\gsim8M_{\odot}$) 
owing to the dearth of direct
observations with the requisite resolution. Consequently, persistent
gaps in our knowledge of the formation process for OB stars have
remained (e.g., Bally \& Zinnecker 2005; Zinnecker \& Yorke 2007; Tan 2008). 

We have used
the National Radio Astronomy Observatory's\footnote{The  
National Radio Astronomy Observatory is a facility of the National
Science Foundation operated under cooperative agreement by
Associated Universities, Inc.} 
Very Long Baseline Array (VLBA) to monitor the SiO maser emission
surrounding \SI\ at monthly intervals over several years
(see also Greenhill et al. 2004b). Our images have higher
dynamic range and higher cadence than any observations of the \SI\
masers to date.
Here we showcase our results in the form of 19-epoch movies
spanning 21 months.
These movies chronicle for the first time 
the kinematics and evolution of the molecular material $\sim$20-100~AU from a 
massive YSO.
This is part of a series of papers examining
\SI\ on 10-1000~AU scales based on
VLBA monitoring and complementary observations of other
maser and thermal lines using the Very Large Array (VLA) and the Green
Bank Telescope
(e.g., Goddi et al. 2009a,b and in preparation; 
Greenhill et al., in preparation; Matthews et al., in preparation). 
Together, these data are allowing us to forge
a comprehensive new picture of \SI\ and its role in shaping the Orion~KL
region.

\section{Observations and Data Reduction\protect\label{observations}}
The $^{28}$SiO $v$=1 and $v$=2, $J$=1-0 transitions toward 
\SI\ ($\alpha_{J2000}$=$05^{\rm h}35^{\rm m}$14.5098$^{\rm s}$,
$\delta_{J2000}$=$-05^{\circ}22'$\as{30}{4820}) were observed using
the VLBA between 2001 March~19 and
2002 December~10. The separation between observations was
approximately one month, although no usable data were obtained during
two months in mid-2002 (Table~1).
Observations were made with the 10 antennas of the VLBA,
together with a single antenna from the VLA.
The $v$=1 and $v$=2 transitions were observed
simultaneously, each in a single polarization (RR)
with a 15.9~MHz
bandwidth centered at 7.5~\kms\ relative to the
Local Standard of Rest (LSR). The adopted rest frequencies for the
$v$=1 and $v$=2 transitions  were 43.122024~GHz and 42.820432~GHz,
respectively. The data were processed with the VLBA correlator,
yielding 512 spectral
channels with a spacing of $\sim$0.2~\kms\ in each of the two IFs.   
Correlator dump 
times were 2.097~s. 
The total integration time on \SI\ during each
epoch (excluding calibration measurements) was $\sim$6 hours for each 
transition.

We performed all data reduction using the Astronomical
Imaging Processing System (AIPS). Because our science goals require
high precision, high dynamic range measurements, we took
particular care to minimize systematic errors in each stage of our
reduction and 
to maintain accurate astrometric registration 
between the $v$=1 and $v$=2 data (which are very sensitive to delay
calibration errors because of the 300~MHz frequency separation between 
transitions). 
To insure uniform processing of each
epoch, we developed a partially automated 
POPS script to facilitate the implementation of 
certain reduction steps. These semi-automated steps were interspersed with
frequent manual checks of the data.

Corrections for errors in the {\it a priori} Earth orientation parameter
and station positions (computed starting with the USNO 2004b geodetic VLBI
solution\footnote{http://rorf.usno.navy.mil/solutions/2004b/}) 
were applied to the data, followed by
digital sampler corrections for the VLBA correlator. If uncorrected,
these position errors will lead to phase and delay errors which
subsequently cannot be calibrated out.

Next, a preliminary amplitude calibration was performed using the default gain
curves and system temperatures provided by NRAO. Subsequently,
an improved amplitude scale and pointing corrections were 
computed based on comparison of a time series of total-power spectra
to a template for each station.

One-time atmospheric delay and instrumental phase and delay offsets
between the IFs were estimated from fringe fits to a one-minute
portion of a calibrator scan (0528+134 or 4C39.25), 
followed by a second
fringe-fit to both calibrators (including all times)
to solve for the residual delays and rates. Solutions for both IFs
were derived independently. However, after ensuring that no systematic
offsets were present between the two sets of solutions, the 
rate solutions from IF2 ($v$=1) were copied and applied to 
IF1 ($v$=2) in order to minimize phase drifts between the two IFs (which may
in turn lead to astrometric misalignments between the two frequencies).  
The time series phase differences for the two IFs for each antenna 
were examined following these steps to verify that no
phase wraps were present during the track. 

The amplitude
portion of the bandpass calibration 
was computed using the total power data for the
calibrators 0528+134 and 4C39.25, and the phase
portion was derived using a complex polynomial fit to the
cross-power data of the same calibrators. 
Time-dependent frequency
shifts were then applied to the line data to compensate for changes caused by
the Earth's motion over the course of the observation. 

To refine the astrometric position of the maser and further constrain
delay-like errors in the calibration that would affect the relative
astrometry of the $v$=1 and $v$=2 emission, a strong,
compact $v$=1 maser reference feature was identified 
in each epoch through examination of cross-power spectra on a variety
of baselines. Each feature selected  (see Table~2) was one that showed
roughly constant visibility amplitudes as  function of baseline length
and orientation.
A fringe fit was performed on this reference feature to determine fringe
rates. Following this, a fringe-rate analysis 
was performed 
to solve for the position of the reference feature relative to the
{\it a priori} correlator position. This analysis was performed
outside of AIPS,  using a non-linear least-squares algorithm to fit 
a sinusoid to the measured
fringe rates as a function of time and simultaneously solve for a single
vertical atmospheric delay error term in the correlator model for each 
station. 
The position offsets determined by this method (Table~2) were then applied to
both IFs. Typical uncertainties in RA and DEC 
were 5~mas and 20~mas, respectively. The proximity of Source~I to
the celestial equator accounts for the larger
uncertainty in the latter coordinate.

Self-calibration was performed on the $v$=1 reference channel, first
in phase only, then in both amplitude and
phase. Typically a total of 
$\sim$25-30 iterations was required. Because the source's position on
the sky compromises the $u$-$v$ coverage, 
care was exercised to
avoid artifacts in the clean component models used for each subsequent
self-calibration iteration. Final solutions were 
applied to the full spectral line data sets for both the $v$=1 and
$v$=2 transitions. Finally, several 
additional iterations of self-calibration were
used to further improve the $v$=2 solutions before imaging the full
line data sets.

Imaging and deconvolution of the fully calibrated SiO data sets were
performed using robust weighting (Briggs 1995) with ${\cal R}$=0. CLEAN boxes
were placed around emission regions 
in each spectral channel individually to
reduce the effects of clean biasing (e.g., Cotton 2007). All 
channels were cleaned to a level of 20~mJy beam$^{-1}$.
This depth of cleaning balanced sidelobe removal against
the production of deconvolution artifacts that may result from 
over-cleaning. These data are particularly susceptible to such
artifacts as a result of a combination of the source's location on the
sky and the limited $u$-$v$ coverage of the array.
The resulting
deconvolved images have a field-of-view
of $\sim$\as{0}{4}, and in all cases we used a restoring beam of
0.55$\times$0.19~mas with a position angle of 0$^{\circ}$.  The 
1$\sigma$ RMS noise in individual channels ranged from
$\sim$8-10~mJy beam$^{-1}$ for line-free channels (consistent with
expected thermal noise) 
to $\sim$9-13~mJy beam$^{-1}$ for channels with strong
line emission.\footnote{These are global values
based on Gaussian fits to histograms of the pixel values across
the entire channel image; the noise may be higher or lower in
certain regions of the image depending on the emission distribution.} 
Our resulting imaging data have  higher dynamic range
(up to $\sim$8000)
than any previously obtained VLBI maps of the \SI\ SiO masers 
and
also recover a significantly larger fraction of the total SiO flux
($70-90$\%; Matthews et al., in preparation). 

Zeroth, first, and second moment maps were computed from
each $v$=1 or $v$=2 data cube (using the AIPS task MOMNT); 
combined moments containing both
transitions were also produced.
Consecutive epochs were registered by maximizing the cross-correlation
between maps of the logarithm of the velocity-integrated intensity.
This method
removes any component of motion caused by a
proper motion of \SI\ (cf. G\'omez et al. 2008; Goddi et al., in preparation).
When registering maps separated in time by roughly one month, this
method appears to be robust to within approximately one pixel (0.05~mas),
although for some epochs, misalignments between the two SiO transitions of 
up to $\sim$0.3~mas may be present as a result of residual calibration
uncertainties. 

\begin{deluxetable}{lclc}
\tabletypesize{\scriptsize}
\tablewidth{0pc}
\tablenum{1}
\tablecaption{Summary of Observations}
\tablehead{
\colhead{Date} & \colhead{MJD} & \colhead{Project
  Code} & \colhead{Flagged}\\
\colhead{(1)} & \colhead{(2)} & \colhead{(3)} & \colhead{(4)}}

\startdata
2001Mar19 & 51987.5 &BG118A & MK\\
2001Apr16 & 52015.5 &BG118B & MK\\
2001May18 & 52047.5 &BG118C & ...\\
2001Jun22 & 52082.5 &BG118D & ... \\
2001Jul22 & 52112.5 &BG118E & ... \\
2001Aug24 & 52145.5 &BG118F & SC,HN \\
2001Sep19 & 52171.5 &BG118G & FD\\
2001Oct21 & 52203.5 &BG118H& ...\\
2001Nov19 & 52232.5 &BG118I & FD\\
2001Dec21 & 52264.5 &BG118J & ...\\
2002Jan27 & 52301.5 &BG118K & ...\\
2002Mar02 & 52335.5&BG118L & MK,Y\\
2002Apr01 & 52365.5& BG118M & ...\\
2002Apr29$^{\dagger}$ & 52393.5 & BG118N & ...\\
2002May27$^{\dagger}$ & 52421.5 & BG118O & ...\\
2002Jun28 & 52453.5 &BG118P & ...\\
2002Aug09 & 52495.5 &BG129A & BR,Y\\
2002Sep05 & 52522.5& BG129B & HN\\
2002Oct07 & 52554.5& BG129C & ...\\
2002Nov06 & 52584.5 &BG129D & ...\\
2002Dec10 & 52618.5& BG129E & ...\\

\enddata

\tablecomments{Explanation of columns: (1) date of observation; (2)
modified Julian date; (3) VLBA project code; (4) stations flagged due to
weather or 
hardware problems: MK (Mauna Kea); SC (St. Croix); HN (Hancock); FD
(Fort Davis); Y (VLA); BR
(Brewster). }
\tablenotetext{\dagger}{Failed observation}
\end{deluxetable}

\begin{deluxetable}{lcrcc}
\tabletypesize{\scriptsize}
\tablewidth{0pc}
\tablenum{2}
\tablecaption{Astrometry of $v$=1 Reference Features}
\tablehead{
\colhead{Project Code} & \colhead{Ref. Chan.} & \colhead{$V_{\rm ref}$ (\kms)} &
\colhead{$\Delta$RA (arcsec)} & \colhead{$\Delta$DEC (arcsec)} \\
\colhead{(1)} & \colhead{(2)} & \colhead{(3)} & \colhead{(4)} & \colhead{(5)}}

\startdata
BG118A & 229 &+13.4 & +0.032 &  +0.015\\
BG118B & 228 &+13.6& $-$0.015   & $-$0.054\\
BG118C & 338 &$-$10.2& +0.087 & $-$0.077 \\
BG118D$^{*}$ & 336 &$-9.7$ & +0.098 &  $-$0.119 \\
BG118E & 336 &$-9.7$ & +0.098 &  $-$0.119\\
BG118F & 333&$-9.1$ & +0.086 &  $-$0.134\\
BG118G & 225 &+14.3& +0.022 &  +0.059 \\
BG118H& 334 &$-9.3$& +0.101 & $-$0.078 \\
BG118I & 333 &$-9.1$ & +0.096 & $-$0.101\\
BG118J$^{\dagger}$ & 334 &$-9.3$ & +0.080  & $-$0.094\\
BG118K & 335 &$-9.5$& +0.082  & $-$0.145\\
BG118L & 198 &+20.1& $-$0.050 & +0.010 \\
BG118M & 338&$-10.2$ & +0.086 & $-$0.055 \\
BG118P & 335 &$-9.5$ & +0.092 & $-$0.061\\
BG129A$^{**}$ &  335 &$-9.5$  &  +0.093 & $-$0.084\\
BG129B & 340&$-10.6$ & +0.095  & $-$0.106 \\
BG129C & 198 &+20.1&  +0.033 & $-$0.025\\
BG129D & 338 &$-10.2$ & +0.092 & $-$0.084  \\
BG129E & 340 &$-10.6$ & +0.089 & $-$0.082\\

\enddata

\tablecomments{The adopted correlator position 
was $\alpha_{J2000}$=$05^{\rm h}35^{\rm m}$14.5098$^{\rm s}$,
$\delta_{J2000}$=$-05^{\circ}22'$\as{30}{4820}. 
Explanation of columns: (1) Observational epoch (see
  Table~1); (2) IF2 (SiO $v$=1) spectral channel corresponding to the
  reference feature used for fringe rate
  analysis and self-calibration; (3) LSR velocity corresponding to the
  adopted reference channel; (4) \& (5) position offsets
  in right ascension and declination, respectively, in arcseconds, 
derived for the reference feature through a fringe rate analysis (see
  \S~\ref{observations}).}
\tablenotetext{*}{No astrometric solution was possible;
  values from BG118E were adopted.}
\tablenotetext{\dagger}{Atmospheric terms were not included in the
 solution.}
\tablenotetext{**}{No astrometric solution was possible; the mean
  of the solutions from
  BG118P and BG129B was adopted.}
\end{deluxetable}

\section{Results\protect\label{results}}
\subsection{The SiO Maser Emission Distribution}
\subsubsection{Total Intensity Images\protect\label{morph}}
\begin{figure}
\scalebox{0.5}{\rotatebox{0}
{\hspace{-4.2cm}\includegraphics{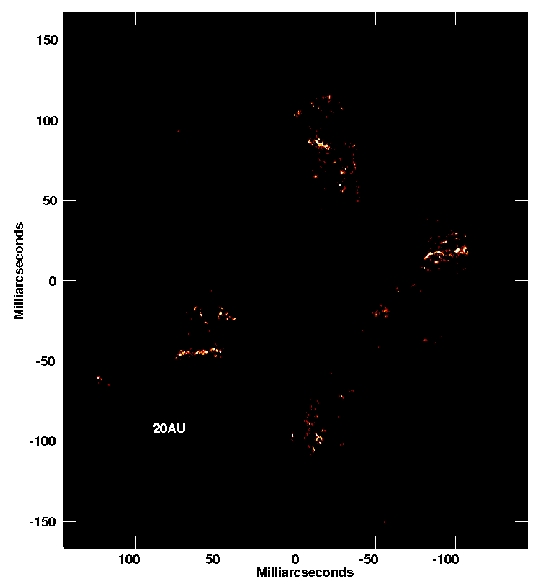}
\hspace{-0.2cm}\includegraphics{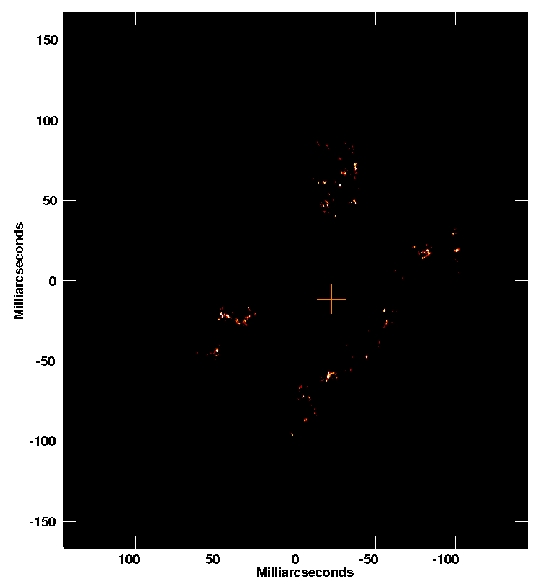}}}
\caption{Velocity-integrated 
total intensity images of the $^{28}$SiO $v$=1, $J$=1-0 
emission (left) and the
$v$=2, $J$=1-0 emission (right) surrounding \SI\, as
observed with the VLBA 
on 2001 March~19. An intensity range of 150 to 5000~Jy beam$^{-1}$ m
s$^{-1}$ is shown using a logarithmic transfer function. The cross in
the right panel
indicates the calculated 
position of the \SI\ radio continuum source  based on the
fringe rate analysis from BG129E (Table~2) and the absolute position and proper
motion measurements of Goddi et al. (in preparation). Note that this
position is slightly displaced from the origin (0,0). }
\protect\label{fig:mom0}
\end{figure}

Total intensity images of the $^{28}$SiO $v$=1 and $v$=2, $J$=1-0
emission toward \SI\ are presented in
Fig.~\ref{fig:mom0}. To guide the reader,
several key features of the maser distribution 
are also illustrated schematically in Fig.~\ref{fig:cartoon}. 

As seen in previous observations of \SI\ with VLBI resolution (Greenhill et
al. 1998; Doeleman et al. 1999; Kim et al. 2008), the bulk
of the SiO emission is located within four ``arms''
of an {\sf X}-shaped pattern, centered on the position
of \SI.  
Along a position angle of
$\sim141^{\circ}$, a $\sim$14-AU-thick dark band with well-defined inner
boundaries is evident, 
inside of which no SiO maser emission is detected. This zone maintains
a constant thickness over many months and harbors the
disk-shaped 7-mm continuum source imaged by Reid et
al. (2007). 

Another key feature of Fig.~\ref{fig:mom0} is the presence
of two ``bridges'' of emission connecting the North and East arms and
the South and West arms, respectively. Bridge emission was detected in
the $v$=1 and/or $v$=2 transitions in all of our observational
epochs. The Western bridge was always
significantly brighter than the Eastern bridge---the latter typically
showing at most a few weak, isolated spots during any given epoch
(e.g., Fig.~\ref{fig:eastbridge}). 
A counterpart to the Western
bridge was first seen in the 1997 VLBA images of Doeleman et al. (2004), and
those authors noted that it could not be readily explained
in the context of existing biconical outflow models for \SI, which assumed
that the symmetry  axis of the outflow lay along the
northwest/southeast direction (cf. Greenhill et al. 1998; Doeleman et
al. 1999). Indeed, this bridge emission is now a key piece of
evidence demanding a new geometric and dynamical model for \SI\
(\S~\ref{globvel} \& 
\ref{model}; see also Greenhill et al. 2004b; Kim et al. 2008).

Comparison of the maps of the two SiO transitions in
Fig.~\ref{fig:mom0} shows that on
average, the $v$=2 emission arises at smaller projected distances from
\SI\ than the $v$=1 emission (see also Kim et al. 2008; Goddi et al. 2009a), 
although there is considerable overlap. 
To better quantify this, in
Fig.~\ref{fig:histo} we plot histograms of the number of discrete maser
features identified in each of 
the two SiO transitions (see \S~\ref{proper}) as a
function of distance from Source~I. To minimize the impact of
projection effects, only spots within the four ``arm''
regions are included. We see that $v$=2 features
dominate at smaller distances from the star, 
while the $v$=1 features become more prevalent at larger
radii. Beyond
$r\gsim$45~AU, 
$v$=2 emission fades, and only $v$=1 features are observed. 
The radial offset, $\Delta r_{p}$, between the locations where the 
peak number densities of $v$=1 and $v$=2
features occur is $\sim$11~AU. 
This is substantially larger than the offsets observed
in evolved stars ($\Delta r_{p}\lsim$1~AU; e.g., Desmurs et al. 2000; 
Soria-Ruiz et
al. 2004; Yi et al. 2005). Furthermore, we find no evidence for
variation in the relative distributions of the $v$=1 and $v$=2 masers during
our monitoring. This too is in contrast to the case of evolved
stars, where the relative position of the two transitions may be seen
to vary as a
function of stellar phase (i.e., on timescales of months; 
see Gray \& Humphreys 2000; Yi et al. 2005). Nonetheless, we cannot
yet exclude that changes in the relative distributions of the $v$=1
and $v$=2 masers around \SI\ may evolve on timescales of several years
or more. 

The presence of an overlap region for the SiO $v$=1 and $v$=2 masers
provides an interesting constraint on the dominant excitation mechanism
for the masers. Purely radiative pumps (e.g., Alcolea 1989; Bujarrabal
1994a,b) do not produce spatial coincidence of the $v$=1 and $v$=2
masers. On the other hand, collisional pumps lead to a large amount of
overlap in the parameter space for these two transitions (e.g.,
Lockett \& Elitzur 1992), in agreement with our \SI\ observations. 

Preliminary radiative transfer modeling by
Goddi et al. (2009a) using a
large velocity gradient code provides some insight into the
physical conditions (temperature, density) required to produce the
observed $v$=1 and $v$=2 maser distributions.
Using a model that included
both collisional
and radiative pumping, as well as physical 
parameters appropriate for the \SI\ region, 
Goddi et al.  found that 
the $^{28}$SiO $v$=2 maser will survive in a stronger, 
hotter radiation field than
the $v$=1 maser and that it will be optimized at higher kinetic
temperatures ($T_{k}>$2000~K) where it can become more strongly
inverted than the $v$=1 transition. This  is consistent with
the observation that 
$v$=2 masers tend to arise closer to \SI\ than the $v$=1 masers.
The Goddi et
al. model also predicts overlap in the 
range of densities where the $v$=1 and $v$=2 transitions
can occur (i.e., $n_{H_{2}}=10^{8}-10^{10}$~cm$^{-3}$ for $v$=1 and 
$n_{H_{2}}=10^{9}-10^{11}$~cm$^{-3}$ 
for $v$=2, respectively). This suggests that the mean gas
densities  within
this overlap zone observed around \SI\ 
should be $n_{H_{2}}\sim10^{9.5}$~cm$^{-3}$. More sophisticated
modelling that treats the effects of maser saturation and line overlap
will be presented by Humphreys et al. (in preparation).

\subsubsection{The Temporal Evolution of the Intensity 
Distribution\protect\label{movie}}
\begin{figure}
\scalebox{1.0}{\rotatebox{0}
{\hspace{-4.7cm}\vspace{-3.0cm}\includegraphics{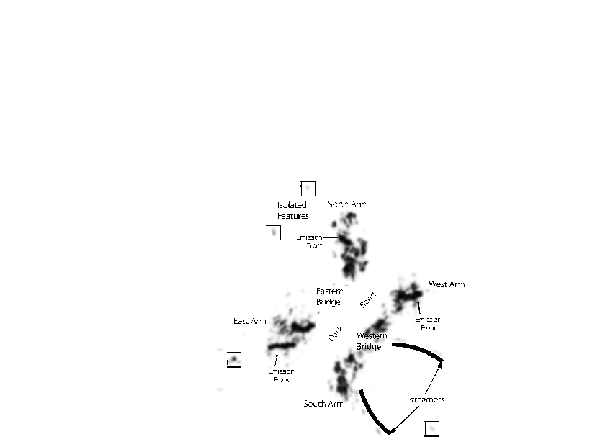}
}}
\caption{Schematic cartoon illustrating several key features of the
  \SI\ maser distribution that are discussed in the Text. }
\protect\label{fig:cartoon}
\end{figure}

\begin{figure}
\hspace{2.5cm}
\vspace{3.5cm}
\scalebox{0.5}{\rotatebox{0}{\includegraphics{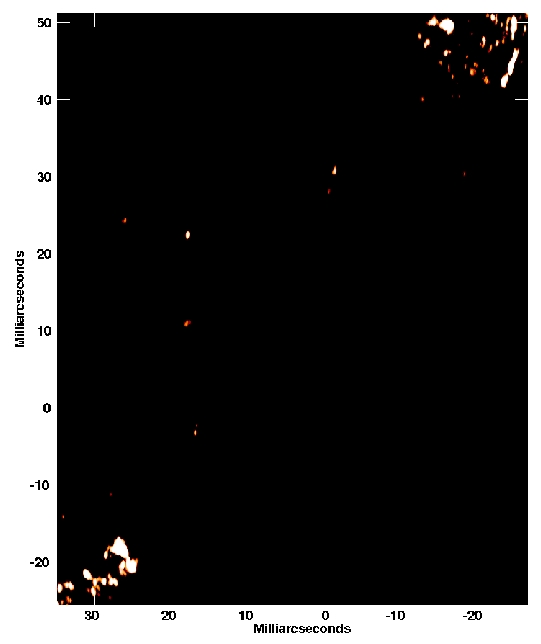}}}
\vspace{-2.0cm}
\caption{Total intensity map showing the combined $^{28}$SiO
$v$=1 and $v$=2, $J$=1-0 emission within the Eastern bridge region
(cf. Fig.~\ref{fig:cartoon}) 
on 2002~June~28. Part of the base of the North arm is visible in the upper
left corner, and the top of the East arm is visible in the lower left.
The intensity range shown is 0 to 500~Jy beam$^{-1}$ m
s$^{-1}$ using a logarithmic transfer function. }
\protect\label{fig:eastbridge}
\end{figure}

\begin{figure}
\scalebox{0.8}{\rotatebox{0}
{\includegraphics{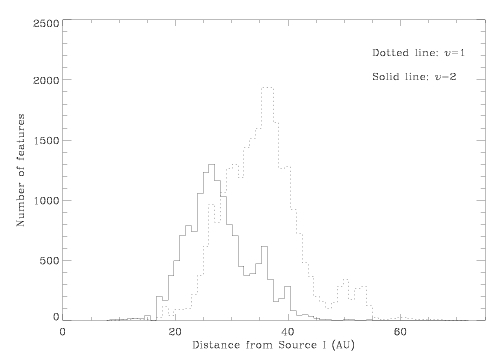}}}
\caption{Histograms showing the total number of discrete maser features
  observed in the $^{28}$SiO $v$=1 transition (dotted line) and the
  $v$=2 transition (solid line) over the course of 19 epochs. 
Discrete features were identified using an automated
  algorithm as described in
  \S~\ref{proper}. Only features arising within the four main arm
  regions (cf. Fig.~\ref{fig:cartoon}) are plotted. Bin sizes are
  1.0~AU. Positions were measured relative to the position of \SI\ as
  indicated on Fig.~\ref{fig:mom0}a.}
\protect\label{fig:histo}
\end{figure}

\begin{figure}
\hspace{0.5cm}
\vspace{4.0cm}
\scalebox{0.8}{\rotatebox{0}{\includegraphics{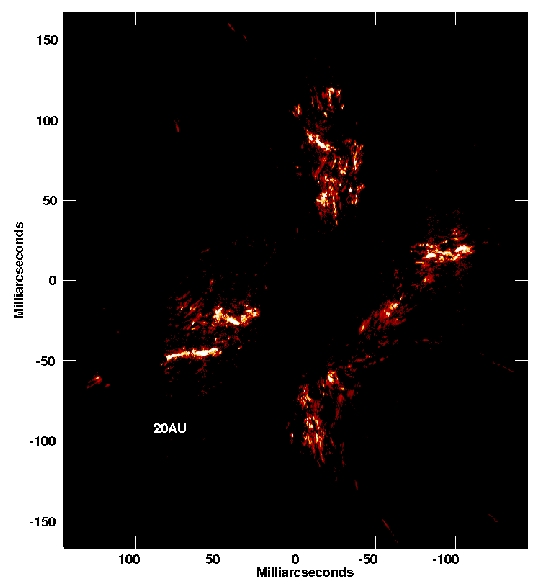}}}
\vspace{-2.0cm}
\caption{Total intensity map showing the combined $^{28}$SiO
$v$=1 and $v$=2, $J$=1-0 emission distribution, summed over 19
observing epochs (see Table~1). An intensity range of 25 to 
30000~Jy beam$^{-1}$ m
s$^{-1}$ is shown using a logarithmic transfer function. A GIF
animation
showing the individual frames comprising this figure is
available at http://www.cfa.harvard.edu/kalypso/Figure5b.gif.}
\protect\label{fig:nest}
\end{figure}

A combined map of both SiO
transitions, summed over all 19 observing epochs (Fig.~\ref{fig:nest}),
reveals the trails of hundreds of individual maser features,
underscoring that at VLBI resolutions, \SI\ 
is the most complex maser source known. Individual maser
features can persist for several months or more,
and a significant fraction 
exhibit detectable motions in the plane of the
sky on timescales of a few weeks; see 
also \S~\ref{proper}). 
In addition to the multitude of maser trails in
Fig.~\ref{fig:nest}, elongated, high surface brightness ``emission fronts'' 
are visible in three of the arms (see also
Fig.~\ref{fig:cartoon}). These fronts are more prominent in the $v$=1
transition (cf. Fig.~\ref{fig:mom0}). 
Analogous features have been seen in 
previous
high-resolution images of the \SI\ SiO masers (Greenhill et al. 1998;
Doeleman et al. 1999, 2004), but our new observations 
show that such structures can remain relatively stable in extent,
position, and velocity over many months. 

An animated movie (Fig.~\ref{fig:nest}, online version) provides a
``4-D'' view 
highlighting the month-to-month changes that
give rise to the intricate structures seen in the 
19-epoch summed image. While
the gross morphology of the maser distribution persists over
the length of the movie, significant temporal changes are evident in
the positions and brightnesses of thousands of individual
features, collectively giving the maser-emitting material
an ``effervescent'' appearance. 

Further inspection reveals that the observed motions
of the maser features exhibit certain systematic
trends. First, outward motions (i.e., away from \SI) are visible along
each of the four arms, as well as in several isolated groups of features
that are detached from the main arms and bridges (see Fig.~\ref{fig:cartoon}).
Radially outward motions are also observed 
within the Western bridge region. However, in addition, maser features along 
the Western bridge 
clearly display a component
of motion {\em tangential} to the bridge. Interestingly, material along
the outer edge of the Western bridge is seen moving to the northwest,
while material along the inner edge moves in the opposite
sense. These patterns most likely arise from emitting material along the rim 
of a rotating, geometrically thick disk whose near side is tipped 
slightly to the
southwest (\S~\ref{model}). Motions within the Eastern bridge also
appear to be consistent with this picture, although features in this
region typically persist for only one or two epochs, making it
more difficult to establish systematic trends.

\begin{figure}
\hspace{0.5cm}
\vspace{3.5cm}
\scalebox{0.8}{\rotatebox{0}{\includegraphics{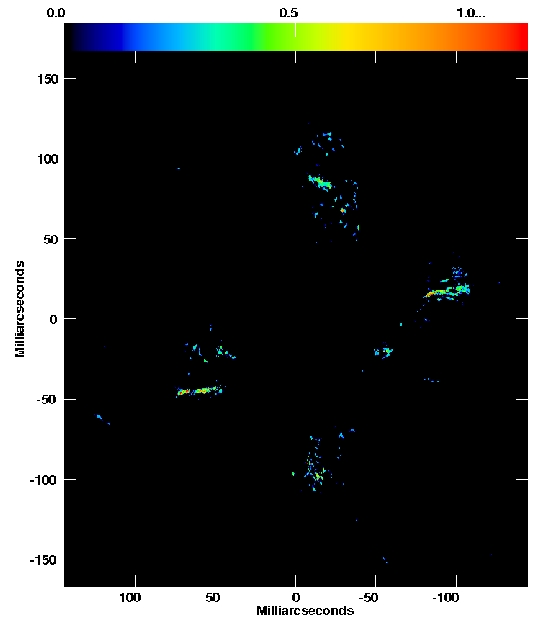}}}
\vspace{-3.0cm}
\caption{Map of the line-of-sight velocity dispersion (in \kms) for the SiO
  $v$=1 transition on 2001 May~18. The color bar is labeled in units
  of \kms. The peak dispersion for this epoch is $\sim$3~\kms.}
\protect\label{fig:veldisp}
\end{figure}

The 
bright, elongated emission fronts described above pose an
interesting contrast to many of 
the other arm features.  Although modest changes in the
position
angle and small-scale morphology 
of these fronts are visible from month to month, their general 
locations and extents remain fairly constant over nearly two
years. We also observe gas within these regions whose 
line-of-sight velocity dispersion
($\sim$1-5~\kms) is several times higher than the mean of the
surrounding material ($\sim$0.3~\kms; Fig.~\ref{fig:veldisp}).

\subsection{The SiO Radial Velocity Field\protect\label{globvel}}
\begin{figure}
\scalebox{0.69}{\rotatebox{-90}{\includegraphics{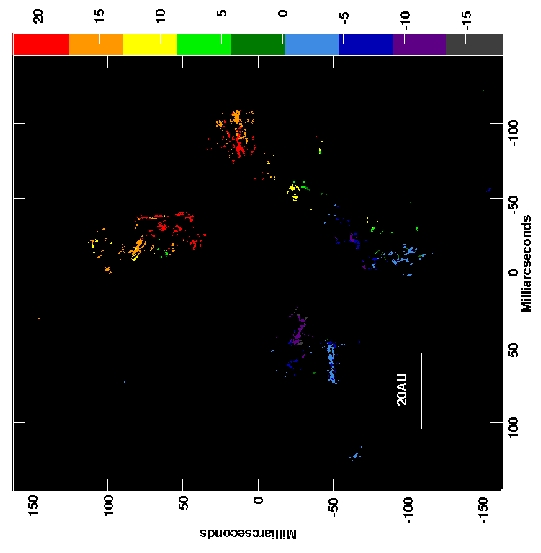}}}
\vspace{2.0cm}
\scalebox{0.56}{\rotatebox{90}{\includegraphics{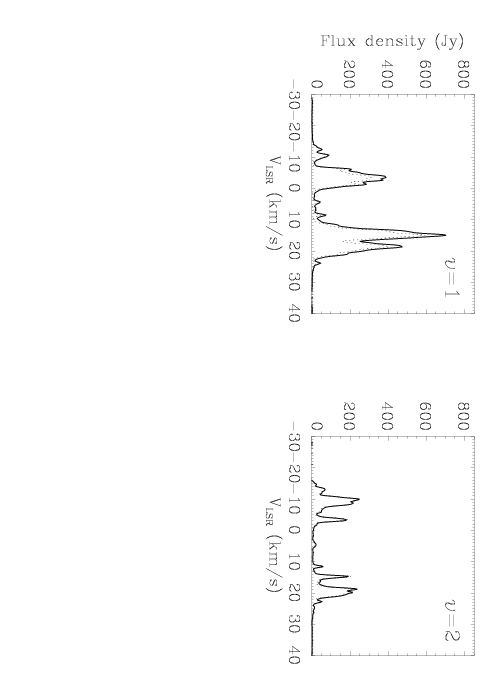}}}
\vspace{-6.0cm}
\caption{Velocity field of the $^{28}$SiO $v$=1 and
$v$=2, $J$=1-0 emission surrounding Orion \SI\ as observed 
on 2001 March~19. The colors indicate measured radial
velocities in \kms. Corresponding integrated spectra are also shown; 
in these panels 
the solid lines show total power spectra and the dashed lines show 
spectra derived from the imaged data. Axes of the spectra 
are LSR velocity in \kms\ and flux density in Jy. The
full velocity spread of the emission varied from epoch to
epoch, but was typically $\sim$40~\kms\ in $v$=1 and $\sim$42~\kms\ in
$v$=2.
A GIF animation of this figure
is available at http://www.cfa.harvard.edu/kalypso/Figure7c.gif.}
\protect\label{fig:mom1}
\end{figure}

In Fig.~\ref{fig:mom1} we show a first moment map 
obtained from the combined SiO $v$=1 and $v$=2 emission on 2001
March~19.  Integrated spectra of the two transitions are also shown. 
The integrated line profiles of 
both the $v$=1 and $v$=2 emission maintained their characteristic
double-peaked shape throughout the course of our
monitoring. A more detailed analysis of the spectral line profiles and
the temporal variability of the SiO masers will be presented by 
Matthews et al. (in preparation). 
Consistent with earlier studies, we also 
see that the redshifted SiO emission arises primarily from the North and
West arms, while the blueshifted emission arises predominantly from 
South and East arms. Furthermore, our new high dynamic range 
images showcase some additional 
characteristics of the SiO distribution 
that were only hinted at by previous investigations.

Fig.~\ref{fig:mom1}  further highlights the
two bridges of emission
connecting the South and West (blue) arms and the East and North (red)
arms, respectively (see also Fig~\ref{fig:mom0} \& \ref{fig:nest}; 
Greenhill et al. 2004b). We see that
emission from both bridges occurs predominantly within a few \kms\ of the
systemic velocity and that a velocity gradient is present in the sense
that the material in the northern end of the bridge has (on average)
higher radial
velocities compared with the material in the southern end. 
We interpret these observations 
as signature of rotational motion about an axis
oriented along the northeast/southwest direction. In
this picture, the red and blue arms represent the receding and
approaching sides, respectively,  of the rotating structure 
(see \S~\ref{model}).

Another notable feature of Fig.~\ref{fig:mom1} is the presence of clear 
systematic velocity gradients along each of the four arms---i.e.,
emission at smaller projected distances from the
source tends to have higher radial velocities than
emission at larger projected radii. The magnitude of the gradients
is $\sim$0.4~\kms~AU$^{-1}$. While there had been some indication of velocity
gradients along
individual arms from earlier imaging studies (Doeleman et al. 2004;
Greenhill et al. 2004b), our new data show similar
gradients along all four arms,
implying that 
they result from a global rather than a localized
phenomenon. This finding has been confirmed independently by Kim et
al. (2008) using data from the VERA array. 
As with the bridge masers, these velocity
gradients are difficult to account for in any model of \SI\ that does
not include rotation about a northeast/southwest axis. 

An animated
version of Fig.~\ref{fig:mom1} (online version) displays the radial
velocity fields derived from all 19 epochs of observations. 
The first moment maps are somewhat noisier than the zeroth moment maps
used to produce the animation in Fig.~\ref{fig:mom0}, leading to
visible north/south artifacts around bright features 
in some frames. Nonetheless,
Fig.~\ref{fig:mom1} helps to highlight the evolution of some of
the smallest and faintest detected features, particularly near the
bridge regions.

\subsection{Proper Motions\protect\label{proper}}
\subsubsection{Methodology}
Fig.~\ref{fig:PM} presents proper motions of individual maser features
derived from our data. Here, positions of the maser features during
each epoch were measured from two-dimensional 
Gaussian fits to the SiO total intensity
images, while radial velocities were taken to be the intensity-weighted mean
velocity at each feature's location. Our typical detection threshold
was $\sim10\sigma$. 

The total numbers of maser spots catalogued over 19 epochs
were 27,857 in the $v$=1
line and 15,448 in the $v$=2 line, respectively.
The resulting $v$=1 and $v$=2
spot catalogues were then systematically searched to identify
features that
persisted over at least {\sl three} epochs (Bridge regions) or {\sl
four} epochs (all other regions) and appeared to move along linear
trajectories in the plane of the sky.

For each of the epochs (beginning with BG118B)
an automated 
search of the preceding epoch was performed
for possible counterparts to each of the identified maser spots.
The search radius was restricted to 1~mas
(corresponding to
$V_{\rm sky}<$35~\kms), with the additional requirement that the
radial velocity change between epochs was less than or equal to 
0.2~\kms. The latter criterion corresponds to
an acceleration $<8\times10^{-3}$~cm s$^{-2}$.

For each
candidate identified, a line was fit between the  position of the
initial spot and the candidate
to delineate
an initial predicted trajectory. The subsequent epoch 
was then searched for spot
candidates lying along the predicted
path to within $\pm$0.15~mas in $x$ and $y$
and $\pm$0.2~\kms\ in $V_{\rm r}$. The former limit corresponds
approximately to our registration uncertainty between epochs, while
the latter is comparable to our velocity resolution. Each time a spot
was found matching these criteria, a linear least
squares fit (in $x$ and $y$) was performed to the resulting set of
three spots to refine the predicted trajectory. Outside of the bridge
regions, this procedure was
then repeated for  the next epoch in the sequence.

Because the region around \SI\ is so crowded with maser features, 
there were a number of
cases where more than one feature from a given epoch was identified as  
a candidate for falling on a particular trajectory.
For these cases, only the trajectory with the lowest 
$\chi^{2}$ value was used to
compute a proper motion vector ($x_{0}$, $y_{0}$, $V_{\rm sky}$, PA) for
the final database. 

The results of the above analysis are summarized in Fig.~\ref{fig:PM}.
Features persisting for more than
3 months (bridge regions) or 4 months (all other regions)
are visible as overlapping arrows. Typical feature 
lifetimes range from less than one month to 3 or 4
months, with occasional features persisting as long as $\sim$11 months.

\subsubsection{Results and Interpretation}
A full catalogue and analysis of the SiO maser proper motions 
will be presented elsewhere (Goddi et al., in preparation), 
but here we describe a few key trends.

The derived 3-D (space) velocities for individual features
range from 5.3~\kms\ to
25.3~\kms, with a mean of 14.0~\kms.  We assume a systemic velocity
for \SI\  
$V_{\rm sys}$=5.5~\kms. We find no evidence for any
systematic decrease in the magnitudes of the motions in the plane of
the sky with increasing
distance from the star, suggesting that the radial velocity gradients 
along the arms (\S~\ref{globvel}) 
are not simply the result of a decelerated outflow, but instead arise
primarily from differential
rotation (see also Kim et al. 2008). 

\begin{figure}
\scalebox{0.8}{\rotatebox{0}{\includegraphics{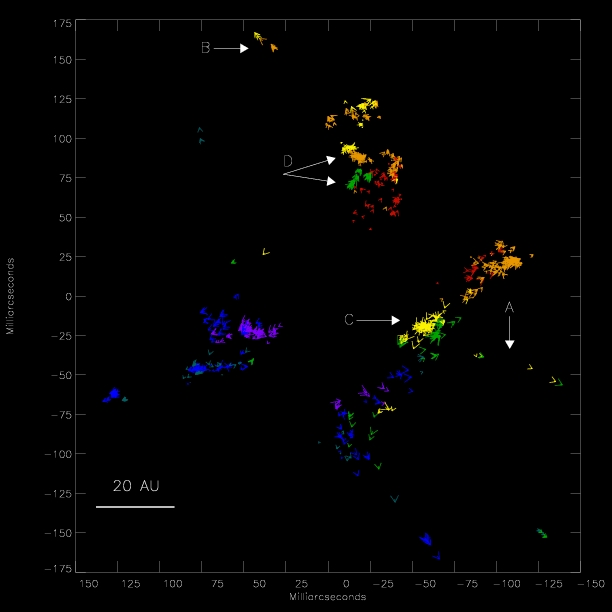}}}
\caption{Proper motions of individual SiO maser features over the course
of 21 months. Both $v$=1 and $v$=2 measurements are shown. 
The {\em color} of
each arrowhead corresponds to its radial velocity (see
Fig.~\ref{fig:mom1} color bar); the {\em size} 
of each arrowhead is proportional to the transverse
velocity (values range from 0.8~\kms\ to 24.0~\kms); the {\em length} 
of each arrow stem indicates the distance
transversed in the plane of the sky over three months (bridge regions)
or four months (all other regions). Some key features described in the
Text (\S~\ref{proper}) are
designated by white letters and arrows.}
\protect\label{fig:PM}
\end{figure}

\begin{figure}
\scalebox{0.6}{\rotatebox{-90}{\includegraphics{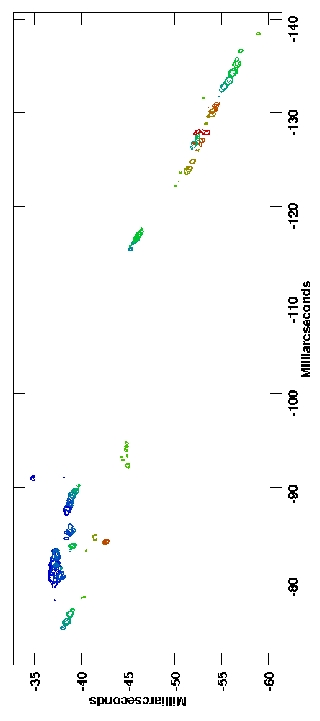}}}
\caption{SiO $v$=1 total intensity contours 
of a ``streamer'' (see
  Fig.~\ref{fig:cartoon} and Fig.~\ref{fig:PM}, feature ``A''). 
Data from 17 epochs are superposed,
  with each epoch contoured in a
  different color. Earlier epochs are shown in blue tones and later
  epochs in red tones.  
  Contour levels are (4,8,...2048)$\times$10~Jy
  m s$^{-1}$. }
\protect\label{fig:streamer}
\end{figure}

\begin{figure}
\hspace{3.0cm}
\scalebox{0.45}{\rotatebox{0}{\includegraphics{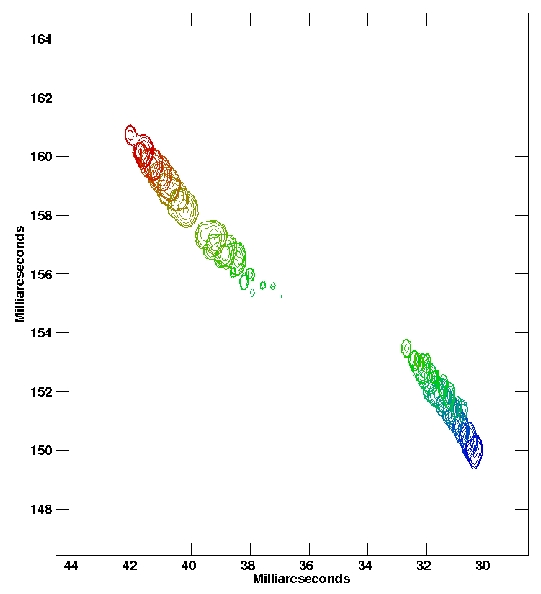}}}
\caption{Similar to Fig.~\ref{fig:streamer}, but for one of the 
``isolated features''
 to the north of \SI\ (see
  Fig.~\ref{fig:cartoon} and Fig.~\ref{fig:PM}, feature ``B''). 
Data from 19 epochs are shown. The gap in the upper feature corresponds
  to a two-month period where no data were obtained (see Table~1). 
Contour levels are (4,8,...2048)$\times$10~Jy
  m s$^{-1}$. }
\protect\label{fig:twizzler}
\end{figure}

The proper motion vectors plotted in Fig.~\ref{fig:PM} 
were derived by conservatively
assuming that the motions were linear over 3-4 months and that the
radial component of the velocity did not change by more than
$\pm$0.2~\kms\ from month to month. 
These initial selection criteria may exclude features with significant 
accelerations. 
A systematic search for accelerations in the spot motions will be
presented in a future paper. However, 
we note that we already see 
qualitative evidence of 
accelerated motions within our data based on the time-integrated
distributions of features along the 
``streamers'' 
and within several 
``isolated'' regions (e.g., the features
labeled ``A'' and ``B'', respectively on Fig.~\ref{fig:PM}). 
Two examples are highlighted in
Fig.~\ref{fig:streamer} \& \ref{fig:twizzler}. 
Although the motions of
{\em individual} maser clumps within the regions shown in
Figs.~\ref{fig:streamer} \& \ref{fig:twizzler} cannot be
distinguished from linear paths to within positional uncertainties, the
time-integrated views reveal the appearance of 
multiple features over multiple epochs that cluster along 
curving arcs, as would be expected if the maser-emitting
material
were being shepherded along twisted magnetic field lines (e.g.,
Banerjee \& Pudritz 2006; see also \S~\ref{MHD}). 

Other noteworthy features of Fig.~\ref{fig:PM} include 
the proper motions along the Western 
bridge (labeled ``C'' on Fig.~\ref{fig:PM}), which exhibit a component
of radially outward motion in addition to two
oppositely directed streams moving tangential to the bridge, as seen
in the animated versions of Fig.~\ref{fig:nest} \& Fig.~\ref{fig:mom1}.
Finally, as indicated by arrow ``D'', in the North arm we find that the 
proper motion vectors  twist by $\sim90^{\circ}$ 
roughly half-way along the arm, again suggestive of material that may
be following non-linear paths.

\section{Discussion\protect\label{discussion}}
\subsection{A New Model for the SiO Maser Kinematics\protect\label{model}}
Earlier VLBI observations of SiO masers around \SI\ established the
{\sf X}-shaped distribution of emission as well as the spatial
separation between the red- and blue-shifted arms. Based on
these observations, Greenhill et al. (1998) and Doeleman et al. (1999)
proposed that the SiO masers arise from
limb-brightened edges of a biconical outflow, oriented along a
southeast/northwest direction. However, our new, more sensitive
observations are better explained by a new model---namely
that the masers are associated with an 
edge-on disk whose rotation axis is oriented along the
northeast/southwest direction (see also Greenhill et al. 2004b; Kim et
al. 2008). Support for this model comes from 
the radial velocity gradients along each of
the four arms (indicative of differential rotation), as well as the
presence of the bridge emission and its associated
velocity gradients (\S~\ref{globvel}). 
Furthermore, the canting of the maser arms (Fig.~\ref{fig:nest})
suggests a reflexive symmetry about a plane whose position angle
(PA$\approx141^{\circ}\pm1^{\circ}$)
closely matches the  ridge of
7-mm continuum emission observed by Reid et al. (2007) on 2000 November~10
(PA$\approx142^{\circ}\pm3^{\circ}$). The presence of the
well-defined dark band in Fig.~\ref{fig:nest} (see 
\S~3.1.1) also suggests that the
conditions favorable for excitation of the SiO masers set in rather abruptly
at a fixed scale height above a flattened, disk-like structure.

Underlying the new kinematic model for \SI\ is the assumption that the
SiO masers trace real, physical motions of
gas clumps rather than, e.g.,  
illumination patterns or shocks transversing a
fixed medium. The contrast between the 
pattern of motions observed in the Western 
bridge versus the arms (\S~\ref{proper})
strongly supports this interpretation, as shocked material would not
be expected to have two apparent kinematic components (both tangential
and radially outward).  Moreover, owing to
gradients in temperature and density between regions close to the
disk plane and the outer reaches of the wind,
shock conditions are unlikely to be similar over the scales of
tens of AU from \SI\ where maser motions are observed.  Finally, the multitude
of linear maser trails seen in Fig.~\ref{fig:nest} also supports a
kinematic interpretation for the SiO masers;
features that move many times their characteristic sizes without
significantly changing morphology 
would not be expected to arise as a shock front
transverses a clumpy, inhomogeneous medium and are a strong indicator
that we are tracing motions of individual clumps. 

\subsection{The Mass of Source I\protect\label{mass}}
One of the longstanding controversies surrounding \SI\ has been the 
mass of the central star. While our present
data can offer important new constraints on this quantity, obtaining a
precise mass estimate is complicated by the likelihood that the
maser-emitting material is not in purely Keplerian
rotation (see also Kim et al. 2008). 
For example, some degree of turbulence is almost certain
present in the gas (e.g., as evidenced by small-scale complexities in the
radial velocity fields in the arm regions), and the outward motions of 
the maser
clumps (\S~\ref{proper}) imply forces acting on the maser gas
opposite to those of gravity (e.g., radiative and/or magnetic
forces).  Such effects can lead to 
underestimates of the enclosed mass 
(e.g., K\"onigl \& Pudritz 2000; Pi\'etu et al. 2005; Bujarrabal et
al. 2005). Moreover, the disk itself might contain up to a few solar
masses of material (Reid et al. 2007) and thus have a non-negligible
mass relative to the central source.

One means of estimating of the mass of \SI\ 
comes from the observed transverse motions of
maser features in the Western bridge region (Fig.~\ref{fig:PM}). 
We assume
that the bridge features lie at $r\sim35$~AU in a circularly rotating
disk (i.e., just outside the edge of ionized inner disk measured by Reid et
al. 2007) and that the \SI\ disk has
a nearly edge-on inclination to our line-of-sight
($i\sim85^{\circ}$). Taking the 
mean space motion of the $v$=2 bridge features ($V_{\rm 3D}\sim$13.5~\kms)
then implies $M_{\star}\gsim7M_{\odot}$.  A second mass estimate can
be derived by
assuming that the material within the four arms is part of an
outflowing wind (see \S~\ref{wind}) and therefore must be moving at or
near escape velocity.  Taking the 
mean space velocities of 
the $v$=2 masers within the four arms ($V_{\rm 3D}\sim16.0$~\kms) and
a fiducial
radius 
$r\sim$25~AU from \SI\  (roughly equal to the radius of the midpoint of the
base of each arm from \SI) also implies a central mass  of
$M_{\star}\gsim7M_{\odot}$.  
These kinematically-derived masses are somewhat
smaller than the
estimate of Reid et al.  (2007) based on the 7-mm radio continuum
luminosity ($M_{\star}\approx10M_{\odot}$). Since a star with
$M_{\star}\lsim7M_{\odot}$ would be unable to produce an \HII\ region
and would not have sufficient luminosity to power the SiO maser
emission ($L_{\star}\gsim10^{4}~L_{\odot}$; Menten \& Reid 1995),
it therefore seems probable that
\SI\ is  somewhat higher than the above kinematically determined
values---i.e., $M_{\star}\sim8$-$10M_{\odot}$. 

If magnetic fields are threading the \SI\ disk and are responsible for
powering the wind (see \S~\ref{MHD}), then the models described by 
K\"onigl \& Pudritz
(2000) predict that  only the material at
the {\sl base} of the wind is expected to exhibit 
Keplerian rotation. We
therefore have 
measured the locations of the peak rotational velocity of the SiO
emission along the northeastern edge
of the dark band that runs parallel to the disk midplane 
(see Fig.~\ref{fig:cartoon}). Taking a mean from the
19 epochs of data and from the
red- and blue-shifted sides of the disk, we find 
$|V_{\rm max}|\approx$19~\kms\ at $r\approx20$~AU. This implies
$M_{\star}\gsim8M_{\odot}$. 

If dust is mixed with the SiO maser-emitting gas as proposed by 
Elitzur (1982), then
radiation pressure on the grains could also influence
the  gas kinematics, 
assuming the grains can efficiently
transfer some of their momentum to the gas. This would again lead to
observed velocities that are smaller than predicted by Kepler's Law
for a given mass.  However, estimating the
magnitude of this effect on the observed gas velocities 
will depend on the intrinsic stellar mass and
luminosity, as well as the detailed properties of the grains---all of
which are uncertain (e.g., Kwok 1975). Moreover, 
it is unclear
whether dust could survive at the temperatures expected near the base
of the \SI\ wind ($T\gsim$2000~K; Goddi et al. 2009a; see also below).

We emphasize that the interpretation of \SI\ as a single, luminous YSO
with $M_{\star}\sim8$-$10M_{\odot}$ 
seems to provide the simplest explanation for {\em both} the radio continuum
observations and the SiO maser kinematics. A binary of equivalent
mass, or a less massive, less luminous star, would not be able to
produce the observed radio continuum emission nor provide the necessary
luminosity to power the SiO masers. On the other hand, a
significantly more massive star would be difficult to reconcile with
the observed SiO kinematics (i.e., the
transverse motions in the bridges, which we interpret as
material orbiting the central mass, and the
systematically outward
motions in the arms, which we interpret as
material traveling at or near escape
velocity). Our results therefore seem to be  in
contradiction with the scenario recently proposed by G\'omez et al. (2008), in
which \SI\ was ejected from a multiple stellar system $\sim$500~years
ago and now comprises a tight binary with a mass in the range
$12~M_{\odot}<M_{\star}<19~M_{\odot}$.

In addition to the problem of the 
discrepancy between our derived mass for Source~I 
and the value proposed by G\'omez et al. (2008), accounting for 
the properties of Source~I's disk in such a picture (e.g.,
its size, density, and symmetry) might also be problematic. 
The passage of another star within $\sim230\pm70$~AU of \SI\ (see
G\'omez et al.) and the
subsequent formation of a tight binary 
most likely would have  disrupted any previously
existing disk around \SI\ (e.g., Moeckel \& Bally 2006), requiring
formation 
of the existing disk structure within the past 500~yr. Nonetheless,
under certain conditions, such rapid disk regrowth might be possible
via either Bondi-Hoyle accretion or the tidal shredding of the
interloper.

In the case of Bondi-Hoyle accretion,
the Bondi radius of an accreting object
of mass $M$
is defined as $r_{\rm B}=GM/v^{2}_{i}$, where $G$ is the gravitational
constant and $v_{i}$ is a characteristic velocity, which may be taken
to be the motion of the object with respect to the surrounding
gas (e.g., Krumholz et al. 2006). 
From Goddi et al. (in prep.), the motion of \SI\ with respect to
the ambient medium is $v_{i}\sim$12~\kms, implying $r_{\rm B}\approx$50~AU for
$M=8M_{\odot}$. Such a radius is comparable to the observed size of the \SI\
disk. 

While the mass of the disk surrounding \SI\ is uncertain (see
Reid et al. 2007), we can roughly 
estimate this quantity, $M_{d}$, by assuming the disk shape is a
flattened cylinder with $r\sim r_{B}=$50~AU and $h$=14~AU (\S~\ref{morph})
and that the mean
particle density is comparable to the value 
required to explain the SiO maser emission
($n\approx10^{10}$~cm$^{-3}$). Assuming the bulk of the disk material
is ionized, we
adopt a mean molecular weight per particle of $\mu$=0.6, implying
$M_{d}\approx0.002M_{\odot}$. Given a time frame of 500~yr, 
this implies a minimum required mass flux 
${\dot M}_{d}\sim 4\times10^{-6}~M_{\odot}$~yr$^{-1}$ (comparable to
the mass {\em outflow} rate estimated from the SiO $v$=0 emission;
Greenhill et al., in prep.). 
Using Equation~1 of Krumholz et al. (2006), we can now estimate the
required ambient density to support this mass accretion
rate as $\rho_{a}={\dot M}_{d}v^{3}_{i}[4\pi
    G^{2}M^{2}_{\star}]^{-1}\approx 3.0\times10^{-17}$~g~cm$^{-3}$.
Assuming the ambient material is purely molecular ($\mu$=2.3) then
implies an ambient particle density $n_{\rm H_{2}}\approx
8\times10^{6}$. Although the latter value is rather high, it
is comparable to values previously measured for the Orion hot
core region (e.g., Masson \& Mundy 1988 and references therein) and
thus may be roughly consistent with plausible values for this Orion~KL
region. We conclude that in
the absence of more sophisticated calculations, we  cannot
rule out disk augmentation or 
rebuilding via Bondi-Hoyle accretion may have occurred during the past
500~yr. However, we note that one additional
caveat is that disks formed in this manner are predicted to have rather
chaotic and 
asymmetric structures 
(e.g., Krumholz et al.), in contrast to the \SI\ case.

One additional scenario that might reconcile our new \SI\
measurements with the findings of G\'omez et al. (2008) is
the possibility that \SI\ has recently formed or altered its 
disk by tidally shredding a lower
mass interloper (see Davies et al. 2006). If this interloper had a
mass as high as $M_{\star}\approx 3M_{\odot}$, this would bring 
the total mass of the \SI\
  system in marginal agreement with values proposed by
  G\'omez et al. Furthermore, such an event might offer a natural
  explanation for the maser emission and outflows associated with
  \SI\ (see Bally \& Zinnecker 2005). 
Nonetheless, this picture would likely require a considerably closer passage
  between the two stars than estimated by G\'omez et al. (i.e., 
as close as a few tens of stellar radii; 
see Davies et al.). In addition, the relatively large resulting 
mass of the disk
may in turn require an implausibly luminous central star in
  order to account for the observed radio continuum emission (see Reid
  et al. 2007). 

In summary, while our latest observations of \SI\ present a compelling case 
for disk-mediated accretion in a massive YSO,
it is clear that further modeling will be
required to explore possible interaction scenarios 
and to better constrain whether such an event is likely to have
influenced its present disk and outflow 
properties.   New N-body simulations as
well as further discussion 
of the interaction history of \SI\ will be presented 
by Goddi et al. (in preparation).

\subsection{What Drives the SiO Maser Emission from 
Source I?\protect\label{wind}}
Because SiO maser emission is extremely rare around YSOs, its origin in the
case of \SI\ has been another longstanding puzzle. 
Cunningham et al. (2005)  proposed
that the \SI\ masers arise from a shear layer along the walls of a cavity
that has been
evacuated as a bipolar wind expands into a rotating, collapsing
envelope. This model is able to reproduce the magnitudes and
directions of the proper
motions in the arms as well as the frothy appearance of the
SiO-emitting 
material. However, the
observed breadth of the arms appears to be greater than expected 
for an interface region. Moreover, this model cannot readily explain several 
other features
of the SiO masers, including the transverse motions observed in 
the inter-arm bridges, the
presence of groups of maser features beyond the four main arms (the
``isolated'' features in Fig.~\ref{fig:cartoon}), or the 
linear trajectories of features that persist unperturbed
over many months (Fig.~\ref{fig:nest}, \ref{fig:streamer}, \& 
\ref{fig:twizzler}). 
The Cunningham et al. model also predicts
higher densities along the outer edges of the arms (their Fig.~1),
which appears to be
inconsistent with our observation that the $v$=2 masers (which preferentially
occur in higher density gas) lie on average closer to \SI\ than the
$v$=1 masers (\S~\ref{morph}; Fig.~\ref{fig:histo}). 

Wright et al. (1995) proposed a slightly different scenario---namely that
the SiO masers arise from material
ablated from the surface of an accretion disk by a wind
or outflow. Because our current observations provide evidence
for the presence of an accretion disk, we now favor some variant of this
``boiling disk'' picture. Constraining the driving 
mechanism for this wind will require 
detailed modeling, but here we comment briefly on the likely
applicability of various classes of disk wind models to the \SI\ case.

\subsubsection{Disk Photoionization}
As discussed by Hollenbach et al. (1994), YSOs hot enough to produce
an \HII\ region are capable of mass-loss via photoevaporation of their
disks as material is heated to temperatures in excess of the local
escape temperature. In the ``weak wind'' case, an ionized flow is
predicted to set in beyond the disk radius, $r_{g}$, 
where the sound speed is roughly equal
to the escape speed. For \SI, if we take the sound speed as
$\sim$11~\kms\ (assuming $T=8000$~K on the surface of the ionized
disk; Reid et al. 2007), a mean mass per particle of
1.13$\times10^{-24}$~g within the ionized disk 
(Hollenbach et al. 1994), and
$M_{\star}\approx8M_{\odot}$, 
this predicts $r_{g}\approx1\times10^{15}$~cm--- roughly a factor of three
larger than what is observed.  In addition to this discrepancy, 
the outflowing material 
is predicted to be mostly ionized, raising the problem of how to
maintain sufficient quantities of dense, molecular material in the
disk wind and how to account
for the SiO maser emission in the bridge regions.

\subsubsection{Line-Driven Winds}
For hot stars, radiation pressure mediated by ultraviolet absorption
line opacity offers another means of powering a
wind. 
Classically, such
winds tend to have velocities too high ($\gsim$400~\kms) 
and densities too low ($\rho<< 10^{-14}$ g cm$^{-3}$) to
readily account for the maser-emitting gas surrounding \SI\ (e.g.,
Lamers et al. 1995).
Drew et al. (1998) showed that the presence
of a rotating, circumstellar disk helps to produce
a slower, denser  wind along the equatorial direction. However, for
$M_{\star}\approx10M_{\odot}$, the characteristic wind
speeds at more oblique angles (i.e., angles consistent with the outward
motions along the arms of \SI) still are an order of
magnitude too high to match those of the \SI\ SiO
masers (see Drew et al.). 
Furthermore, the high poloidal velocities predicted by this
model are inconsistent with outflow speeds
inferred from other line tracers, such as SiO $v$=0 emission (Goddi et
al. 2009a; Greenhill et al., in preparation). 
It thus appears that line-mediated radiation pressure is unlikely to play a
significant role in powering the outward migrations of the SiO
maser-emitting gas around \SI.

\subsubsection{Radiatively-Driven (Dust-Mediated) Winds}
Elitzur (1982) has suggested that gas and dust may be mixed within the
zone from which the SiO maser emission arises around \SI. This would
be in
contrast to the situation in the extended atmospheres of evolved
stars, where the SiO maser emission arises from a region just inside
the dust-formation radius (e.g., Reid \& Menten 1997). Elitzur further
showed that in such a case, dust-mediated 
radiation pressure  alone might be able to
account for the outflow of material from \SI. 

Although we cannot yet exclude this general class of model for
driving the \SI\ wind, several of the previous assumptions made by Elitzur
(1982) require revision as a result of more recent,
high-resolution observations. For example, Elitzur assumed a
luminosity for the \SI\ of $10^{5}L_{\odot}$, which is likely an 
order of magnitude too high (Reid et al. 2007) and a mass-loss rate
of $10^{-3}M_{\odot}$~yr$^{-1}$---two orders of magnitude larger
than the value 
recently derived by Greenhill et al. (in preparation) based on SiO
$v$=0 measurements. Elitzur's model
also depends upon energy input by turbulent motions of $\sim$7~\kms\
in order to maintain the gas at a higher temperature than the dust (as
otherwise the gas would rapidly thermalize to the dust blackbody
temperature and population inversion could not be maintained). 
Such large turbulent motions are
inconsistent with the low mean line-of-sight velocity dispersion
of the maser-emitting material seen in our present data 
(Fig.~\ref{fig:veldisp}) as well as
with the ordered velocity field of the SiO masers 
(Fig.~\ref{fig:mom1}). Lastly, as already noted
above, it would likely be difficult for grains to survive within
$\lsim$100~AU from \SI, particularly near the
base of the wind, where temperatures may be $\gsim$2000~K (Goddi et
al. 2009a). Given these new developments, the dust-driven wind
scenario for \SI\ now faces a number of challenges. 

\subsubsection{Magnetohydrodynamic Winds\protect\label{MHD}}
Apparent curved and helical trajectories of certain SiO maser features
(\S\ref{morph} \& \ref{proper}), strongly hint that magnetic 
fields may play a role in shaping the dynamics of the \SI\ region.  
For example, it would seem difficult to explain features such as 
the streamer emulating from the Western bridge 
(Fig.~\ref{fig:streamer}) in the absence of magnetic fields. While
pressure gradients within the disk 
might act to bend the paths of outflowing material, the observations
of pronounced curvature at large vertical displacements from the
plane would
require that the pressure scale height of the disk is comparable to the
observed vertical extent of the masers.  Moreover, we see numerous 
proper motion trajectories that appear to be linear, including 
along the outer edges of
the individual arms, contrary to what would be expected if pressure
gradients were important. 
We therefore suggest that the streamers are more likely 
to be comprised of gas clumps constrained to move 
along a magnetic field lines like ``beads on a wire'' (see e.g., Blandford \&
Payne 1982). 

Previous evidence for a magnetic field associated with \SI\ was
provided by polarization measurements of the
SiO $v=0$, $J$=1-0 and $J$=2-1 emission (Tsuboi et al. 1996; Plambeck et
al. 2003). After correcting for Faraday rotation, the position angle
for the polarization vectors derived by Plambeck et
al. (57$^{\circ}$) is
consistent with magnetic field lines threading roughly perpendicular
to the disk defined by the SiO masers and 7-mm continuum emission
and parallel to the outflow direction traced by the
SiO $v=0$ emission and the H$_{2}$O masers (see Goddi et al. 2009a;
Greenhill et al., in preparation). 

Assuming a magnetic field is present,
a magnetocentrifugal wind 
(e.g., Blandford \& Payne 1982; K\"onigl \& Pudritz 2000) would be
a natural candidate for powering a disk wind from \SI. 
In this scenario, gas clumps 
fragment from the disk, are swept outward along field lines
by hydromagnetic forces (e.g., Emmering et al. 1992), and are
induced to excite maser emission 
when irradiated, shocked, or heated by collisions. 
Magnetic phenomena may also provide an explanation for the elongated
emission fronts visible in three of the arms. For example, 
tangled magnetic field
lines (e.g., Kigure \& Shibata 2005; Banerjee \& Pudritz 2006)
or instabilities within
a magnetically-driven flow (e.g., Kim \& Ostriker 2000) might account for 
the elevated velocity dispersions and rope-like 
morphologies of these features (see \S~\ref{morph}). Doeleman et al. (1999)
originally proposed that these elongated structures might arise at the shocked
interfaces of an outflow. However, the observation that the emission
fronts are not preferentially oriented 
perpendicular to the outflow direction (as expected for shock
fronts) seems to argue against
this interpretation. 

It is believed that
a necessary condition for launching a magnetically-powered wind is
that the vertical magnetic field is close to equipartition (e.g., Ferreira
2007)---i.e.,
$\frac{1}{2}nV^{2}=\frac{1}{2}B^{2}\mu^{-1}_{0}$, where $n$ is the gas
number density, $V$ is the mean velocity of a gas molecule, $B$ is the
magnetic field strength, and $\mu_{0}$ is the permeability of free
space. Assuming $n=10^{10}$~cm$^{-3}$ (\S~3.1.1) 
and $V$=14.0~\kms\ (\S~\ref{proper}),
the implied magnetic field strength for
\SI\ is $\sim$0.3~G. One possible source for this field might be
the original interstellar magnetic field threading the molecular cloud out of
which \SI\ was born. Based on  measurements of OH 1665-MHz
masers across a $\sim10^{4}$~AU region surrounding \SI,
Cohen et al. (2006) derived field strengths of 1.8 to 16.3~mG. Since the OH
masers are believed to arise from material with molecular hydrogen
density $n_{\rm H_{2}}\sim 7\times10^{6}$~cm$^{-3}$ (Gray et al. 1992), and
magnetic field strength is expected to
scale as the square root of the gas density, it is 
plausible that field strengths of order the  value predicted by
assuming equipartition might now
be present within the denser, SiO-emitting gas. 

Direct measurements of the magnetic field strength in the disk of
\SI\ would be 
of considerable interest, both for understanding its role during the
accretion/outflow process and for providing new clues on the 
nature of magnetic fields in B-type stars during later evolutionary stages.
Little is presently known about the range of
magnetic field strengths present in  early-type B stars on the main sequence
owing to the difficulty of measuring fields $\lsim$3~kG in rapidly
rotating stars (e.g., Landstreet 1992; Schnerr et
al. 2008 and references therein). Moreover, it has been suggested that
magnetic fields with strengths as low as a few Gauss and originating at the
time of the star's formation might account for the origin of magnetized
neutron stars (Ferrario \& Wickramasinghe 2005). 

We note that even if magnetohydrodynamic forces are not the main
driving mechanism for the \SI\ wind, magnetic forces may still play a
key role in shaping the outflow, similar to what has been proposed for
planetary nebulae (e.g., Blackman 2008 and references therein). 
In addition, magnetic fields may
provide a mechanism to enhance the coupling between gas and dust
grains, thereby increasing the efficiency of a possible dust-driven
wind (Hartquist \& Havnes 1992).

\section{Summary}
We have presented two multi-epoch movies chronicling the evolution of
the $^{28}$SiO $v$=1 and $v$=2, $J$=1-0 maser emission surrounding 
Orion \SI\ over nearly two years. These movies comprise VLBA
observations with $\sim$0.5~mas (0.2~AU) resolution and provide the most
detailed view ever obtained of the dynamics and temporal evolution of
molecular material within $\sim$20-100~AU of a massive YSO. We
interpret the SiO masers surrounding \SI\ as arising from a
wide-angle, bipolar wind
that emanates from a rotating accretion disk viewed nearly
edge-on. We find evidence to support the suggestion 
that magnetic fields are playing a role in
shaping and/or powering this wind. 
The maser kinematics and proper motions, coupled with
constraints from previous radio continuum observations,
imply a mass for the central star of
$\gsim$8$M_{\odot}$. Our study provides compelling
evidence that disk-mediated accretion and low-velocity, 
wide-angle winds are both key elements in the evolution of young stars in
this mass range. However, we cannot exclude the possibility that a
recent encounter has also played a role in shaping the properties of
the \SI\ disk and outflow.

\acknowledgements
We thank Mark Reid for 
valuable technical discussions and for supplying his fit$\underline~$rates
code. We also acknowledge helpful discussions with Mark Krumholz and
comments from our anonymous referee that have helped to improve our presentation.
This project has been supported by NSF grant 0507478 
and a Visiting Scientist appointment to LDM
from the Smithsonian Astrophysical Observatory. 
The data presented here were part of NRAO programs BG118 and BG129.

\clearpage

\end{document}